# Serious Play to Encourage Socialization between Unfamiliar Children Facilitated by a LEGO® Robot

**Nicklas Lind, Nilan Paramarajah, Timothy Merritt**

Aalborg University Aalborg, Denmark
merritt@cs.aau.dk

**ABSTRACT**

Socialization is an essential development skill for preschool children. In collaboration with the LEGO Group, we developed Robert Robot, a simplified robot which enables socialization between children and facilitates shared experiences when meeting for the first time. An exploratory study to observe socialization between preschool children was conducted with 30 respondents in pairs. Additionally, observational data from 212 play sessions with four Robert Robots in the wild were collected. Subsequent analysis found that children have fun as Robert Robot breaks the ice between unfamiliar children. The children relayed audio cues related to the imaginative world of Robert Robot's personalities and mimicked each other as a method of initiating social play and communication with their unfamiliar peers. Furthermore, the study contributes four implications for the design of robots for socialization between children. This chapter provides an example case of serious storytelling using playful interactions engaging children with the character of the robot and the mini narratives around the build requests.

## 1 INTRODUCTION

Play is a vital part of childhood development, especially in the preschool years, where children gain essential skills used throughout their lives, such as motor skills, number and color recognition skills, and social skills [25,48]. Successfully developing socialization skills in the preschool years is key to success in many aspects of life, as it is essential for building friendships and other social relations throughout life [1,25]. Surprisingly, research in the field of child-robot interaction (cHRI) to support the facilitation of social play between unfamiliar as well as familiar play partners is somewhat limited. This shortcoming presents a research opportunity to broaden the understanding of how a simplified social robot might facilitate socialization between preschool children. We identified a need for interactive technologies to support children's development, with a particular focus on supporting play and interpersonal relationships and making it easy and fun for the children to engage with one another at a young age. It is these childhood transition years that it becomes especially crucial for the children to develop their social skills to form meaningful short-term and long-term relations with their peers [1].

We present *Robert Robot*, an interactive anthropomorphic robot designed in collaboration with The LEGO Group to facilitate creative social play for preschool children [56,57].

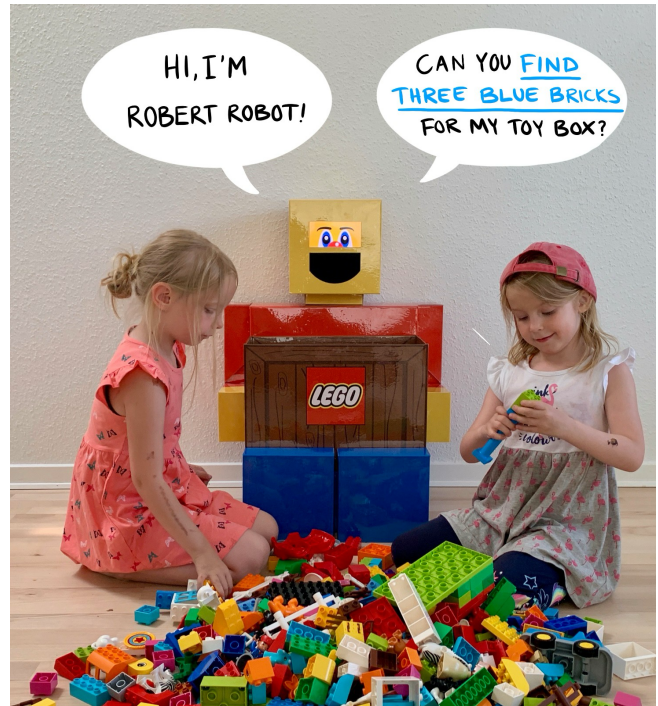

**Figure 1. Two preschool children playing with Robert Robot.**

The play revolves around LEGO® DUPLO® bricks as the primary play material, which has the inherent quality of aiding the children's color, number, and motor skills. Functioning as a research platform for studying child socialization and play behavior, Robert Robot is implemented in a high-fidelity autonomously interactive prototype, which comes to life through three interchangeable and distinct characters with playful personalities.

An exploratory study was designed to collect empirical observations of Robert Robot in use by preschool children in two concurrent tracks: a focused study of 30 preschool children in a controlled play setting, and public observations of 212 play sessions in the wild at the LEGO World 2019 exposition. This provides and in-depth case related to serious storytelling [60] and the results of the study are intended to inform and deepen the understanding of child-robot-interaction, with a particular focus on the social play patterns emerging between unfamiliar preschool children. Following our analysis, we present four implications for the design of simplified robots for facilitating play for preschool children.

## 2 RELATED WORK

In this section, related work that informs this exploratory study is presented with a short overview of the theory and findings concerning child development and play, and child-robot interaction. The related work guides our understanding of play, socialization between children in the preschool years, how researchers observe social play, and children's interactions with social robots.

### 2.1 Child Development and Play

Play has essential intrinsic value as part of children's natural development and learning [1,36,44,52]. As pointed out by Vandenberg, "*play*" is challenging to define [47]. Caillois describes two types of play, with *ludus* (games) being structured activities with explicit rules, and *paidia* (play) as unstructured and spontaneous activities. Games and play emerge when elements from four subcategories, *agon*, *alea*, *mimicry*, and *ilinx*, are combined [13].

Through play, children learn essential life skills, such as motor skills, recognition skills, and social skills [44]. Social competence has long been recognized as paramount in early childhood development [26, 31], as childhood social competence has a marked effect on the subsequent young adult life of the children [48].

Starting around the second year of life, children begin development of mutual peer interaction skills [1,11]. Brownell explored whether children who are beginning to develop their peer interaction skills can adjust their social behavior to the age of their play partners and found that both age and gender mixes influence the social behavior of preschool children [11]. They particularly highlight the positive effects of mixed-gender play as skills tied to the development of social-cognitive abilities. When observing toddlers and preschoolers play, Bailey et al. found that age differences influence the engagement and socialization of the children [4,20].

Previous research has explored how interactive designs can support different levels of social play and how these levels emerge in the play [7]. Beukering et al. developed an open-ended play environment which supported three stages of play, inspired by Parten's seminal work [7]. They found that children first approach the play environment in a solitary manner, which then evolves to parallel play before transforming into group play.

Frameworks for observation of play have been proposed since the early 1900s. Building on two play hierarchies, the socially oriented *Stages of Play* by Parten [32] and the cognition-oriented play types described by Piaget [34], Rubin's *Play Observation Scale* proposes three types of social play: *solitary play*, *parallel play* and *group play* [38]. In solitary play, a child plays apart from other playing children with a distance greater than approximately one meter. In parallel play, the child is attentive of others, but plays independently, often within the one-meter boundary of other children. In group play, the play with other children has a common purpose or goal. The children may strive to attain a competitive goal, but goals are primarily group centered. Rubin outlines four types of cognitive play; *functional play*, *constructive play*, *dramatic play*, and *games-with-rules*. Additionally, Rubin outlines eight non-play behaviors, among them *exploratory*, *unoccupied*, *transition*, *aggression*, and *onlooker* behavior [38]. Narrative and serious storytelling techniques are often used to guide learning activities with children facilitated with digital technologies. An example of research aimed at teaching children about sustainable energy utilized simple storylines and constructive activities (building a house and windmill) with animations on electrochromic displays giving immediate visual feedback that helped reinforce their understanding of the physics of energy [58]. Explorative play has been used to understand how children respond to new technologies—one such example used the simple narrative of designing the patterns on butterfly wings and then moving the butterfly to fly a swarm of 10 drones [59]. In that research, various play patterns emerged including performative social play.

### 2.2 Child-Robot Interaction

Winthrop & McGivney note that as the world changes, children need to learn how to develop skills that will help them succeed in this ever-changing environment [53]. One such change might be a world supported by social robots in many aspects of life, including socialization and play. Social robots are often designed to communicate, express emotions, and maintain relationships [12,18,26].

The design of social robots often implements anthropomorphic features, like eye gaze, and audio cues to express emotion [15,16,46]. The pioneering Kismet [8] is a prime example of a social robot which leverages anthropomorphism to mimic human expressions to elicit a response from users. Kismet allowed its users to engage with it in a face-to-face manner while exchanging different social cues like eye contact and joint attention.

Minimalist, or minimally designed social robots have been explored [28], most notably with Keepon [23,24], an interactive anthropomorphic robot developed for education, entertainment and research, particularly research into therapy for children on the autism spectrum. Kozima et al. found that by having the robot exhibit its attention and emotions in a simple and comprehensible way, children with interpersonal communication difficulties were able to understand the social aspects of the interaction with the robot without becoming overwhelmed or bored [24].

Programmable robots, such as LEGO® Mindstorm, have gained success the marketplace and in the broader education space, supporting the teacher's role in teaching children about complex concepts, such as computer programming [22,30,33].

The cHRI research field has focused on how children attribute agency and intelligence to social robots [5,14,50]. In support of this, Salvatore et al. showcased practical approaches for recognizing engagement between children and social robots [3].

Research has explored how the children's engagement with social robots change over time [51] and found that robots that adapt to children's affective states can sustain the children's engagement with the robots [2,8]. Sabanovic et al. highlight the importance of analyzing social robots in concrete social contexts [29,40], underlining one of the challenges in cHRI for the design of how social robots understand their users and how the users understand the robots underlining one of the critical challenges [6].

## 3 RESEARCH PROBLEM

The current research lacks answers to essential questions regarding children's socialization while playing with robots and how the robots might support the children in forming meaningful short-term and long-term bonds with one another. It is particularly interesting to note that given the vast amount of research done into child-robot interaction and social robots, with much of the related work revolving around developmental scenarios, there still has not been a deep focus on play as the primary activity children engage in with robots, though play is essential for socialization during childhood development.

What potentials are there for a minimalist robot play partner to help break down social barriers and facilitate socialization between preschool children? How do preschool children who do not know each other engage in social play with a minimalist play partner? Moreover, how does their play experiences differ from that of children who do know each other? These are some of the questions that guided this exploratory research study.

## 4 CONCEPTUAL DESIGN

For our study, we designed and built a high-fidelity prototype, self-referentially dubbed *Robert Robot*, a simplified interactive anthropomorphic LEGO® DUPLO® robot which facilitates creative and social play between preschool children. It does so by asking the children for help. With various voice lines, Robert Robot asks the children to help him fill his "toy box". Robert Robot might ask: "*Can you find three blue bricks?*" and, once bricks land in his toy box, he might respond, "*Fantastic! That was just what I needed!*".

Robert Robot's boxy exterior is made from colorful, durable laminated cardboard, borrowing design elements from the LEGO® design aesthetic to express playfulness and child-friendliness. The child-size Robert Robot prototype sits 75 cm tall and is stationary with a storage toy box resting on his legs between his arms (see figure 2). Robert Robot comes to life through an interplay between the physical exterior, simple electronics and an accompanying smartphone app, which doubles as Robert Robot's animated eyes and simultaneously controls all aspects of the interaction with Robert Robot. Robert Robot consists of few electronic components: a smartphone, a Bluetooth speaker secured behind a speaker grill which doubles as the mouth, a contact microphone hidden inside a compartment in the toy box and a powerbank to power the entire setup. Robert Robot is easy to assemble, disassemble, and empty of bricks, and once set up, human intervention is only needed when emptying the toy box.

### 4.1 The Three Robert Robot Characters

In the Robert Robot app, facilitators can choose between three different characters with distinct personalities based on three stylized themes: *Robert Robot*, a whimsical, excitable, and curious personality, *Captain Brick*, a lively pirate, or *Robert Balloon*, a humorous circus clown (figure 3).

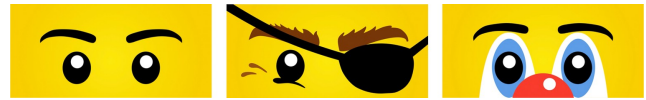

**Figure 3. Still frames of the three characters' eyes.**

### 4.2 Auditory Requests and Feedback Cues

The essential components of Robert Robot's play facilitation are request and feedback voice lines which Robert Robot says to the children in various forms to help filling his toy box. Each character has 50 custom voice lines: 29 request voice lines, 15 feedback voice lines, and 6 idle and personality trait voice lines. Requests are split into three categories: *collect requests*, *build requests,* and *pretend requests*. In the same manner, feedback is split into three categories: *baffled*, *neutral,* and *enthusiastic*.

The voice lines are intended to stimulate the children's developmental skills; they are centered around essential and distinct cognitive themes necessary during childhood development, such as color recognition, number recognition and simple imaginative challenges [1]. The feedback lines are designed to always be neutral or positive to limit negative experiences. Robert Robot's feedback is designed to be humorous, imaginative, and emotional to express the lively personalities. The tone of Robert Robot's voice is light-hearted, extroverted and its vocabulary is kept basic to convey tasks in a language that is easy for children to understand [35]. In many instances, Robert Robot's voice is supported with sound effects to make the robot more childlike and cartoonish, for example: "*Can you find an animal which says [a goat bleating]?*".

### 4.3 The Robert Robot App

The Robert Robot app is essentially the robot's "brain". It controls all the logic behind its weak artificial intelligence [39] which powers all the interactions with the robot. The logic is based on randomness and timed intervals, and determines what requests and feedback voice lines to play in random order, akin to shuffle mode in music playing apps.

### 4.4 Visually Animated Eye Cues

Robert Robot's eyes animate when bricks are registered in the toy box. Cartoonishly designed with stylistic references to the LEGO minifigures [45], the eyes are animated based on basic principles of animation to create an illusion of liveliness [21]. Robert Robot's eyes change dynamically in multiple ways ranging from human-like blinking eyes to heart-shaped eyes. The first and last frame of each animation is identical, which creates a seamlessly transitions from one animation to another. Each character has bespoke animations while some animations recur across all characters. In total, we have designed 48 distinct eye animations; 16 clown-themed animations, 15 pirate-themed animations, and 17 animating Robert Robot eyes.

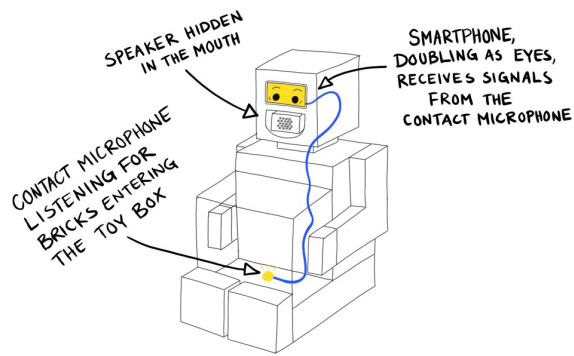

**Figure 2. Illustration of Robert Robot's internal components.**

### 4.5 Play Scenario Example

When initiating a play session, Robert Robot will greet children with an introductory line saying "*Hello! My name is Robert! What's your name? I just love to play with LEGO! Will you help me collect some LEGO bricks for my toy box?*". Once children place bricks into his toy box, the contact microphone will sense it by registering changes in decibel levels when detecting non-airborne vibrations. The input from the contact microphone triggers the app to display an eye animation and keenly respond with a random feedback line through the connected speaker after which a new random request will be played. This all happens in a continuous loop until manually stopped by a facilitator. The frequency of the collect, build, and pretend requests played varies as they are weighted differently in the system by design; collect requests are played 2/4 of the time, while build and pretend are each played 1/4 of the time. Since the underlying system is based on randomness, an activation of Robert Robot can result in approximately 19.500 combinations of feedback, request and eye animations across the three characters, and therefore, any play experience, in terms of the sequence of these cues, is likely unique.

## 5 METHOD

### 5.1 Study Setup

To explore how preschool children socialize when playing with Robert Robot, we designed a two-part study conducted at the LEGO World 2019 exposition in Copenhagen, Denmark. The study ran for four consecutive days. Through our collaboration with The LEGO Group, we were invited to do studies in both the *public exposition area*, as well as a *secluded test area* closed off from the public. We set up a controlled play setting with one Robert Robot in the test area and did concurrent observations of children and their guardians playing with four Robert Robots in public.

### 5.2 Data Collection

We collected a combination of qualitative and quantitative data throughout the study. We primarily focused on the data collected in the test area, while our observations and logged data from the public area was collected to broaden our perspective on our findings in the test area [3,40]. In the test area, we observed and took notes, as well as utilized questionnaires for the guardians as our primary source of data for future analysis. We assigned a volunteer researcher to aid in our data collection throughout the entire study.

Based on preliminary research, we developed a five-point Likert scale questionnaire inspired by the Play Activity Questionnaire [17] to understand the children's expressions of the experience with a guardian's assistance. Additionally, preliminary interviews with the children and their guardian(s) were conducted, as well as debriefings after each play session. While the data collection relied on the guardians as intermediaries, as they have intimate knowledge about their children, simple interview guides with visual artifacts were used to ease the communication with the children when needed [19].

In the Robert Robot app, timestamped log files were continuously monitoring a range of events. Triggering the contact microphone creates one activation event. Logged events include numbers of activations, played requests, the specific feedback following each request, and the total number of feedbacks given to each request. Additionally, the amount of bricks put into the toy box was weighed after each play session.

Observational sheets based on preliminary research were designed to ease the capture of observational data in the hectic exposition environment while ensuring a baseline of consistency across note takers. Each sheet was intended for the observation of four play sessions. We leveraged the Play Observation Scale [38] and templates for f-formations [27] for added context, as well as check boxes for pre-written possible scenarios, identified through our preliminary research and inspired by work on assessing engagement with robots [37,41]. Additionally, duration of play was noted.

### 5.3 Controlled Play Setting

A controlled play setting with one Robert Robot unit was set up to ensure a similar context for studying social behavior and play across the dyads and triads of respondents. The closed-off play setting loosely resembled a lounge area and took up an approximately six square meter area with play mats on the floor for children to play on, while guardians could retreat to observe the play from a table at the opposite end of the room.

### 5.4 Procedure

Each child and their accompanying guardian(s) were briefed separately from their playmate before being introduced to Robert Robot and each other in the play setting. Each play session was allotted around 20 minutes for play, with a subsequent debrief for 5-10 minutes. In the 20 minutes of free play, the children experienced all three of Robert Robot's characters with minimal intervention from a facilitator on standby. Guardians were instructed to fill out questionnaires during or after each play session. These questionnaires have been used to analyze our results. The robot was switched on out of view from the children, after which they continued in open play.

### 5.5 Respondents

We recruited 30 children (19 boys and 11 girls) and their accompanying guardians. We excluded 3 participants from our dataset as they either played alone or primarily with their guardians. In total, we had 5 pairs of familiar children and 7 pairs of unfamiliar children who did not know each other. Additionally, we had one instance where three children, 1 girl and 2 boys, who did not know each other, played together. Participants were aged between 2 and 7 years old (M = 4.26, SD = 1.13).

# 6 FINDINGS

In these sections, we present the quantitative and qualitative findings from our study in the test area and the public area, respectively. The consensus among our participants is that playing with Robert Robot is fun. Notably, unfamiliar children have approximately the same amount of fun as familiar children. Another primary result of the study is that Robert Robot breaks the ice between children who did not know each other before the play session, while still being fun to play with for children who do know each other. Children with an age difference less than two years had more fun and were reportedly happier compared to children with an age difference of two years, echoing findings by Bailey et al. [4].

In total, all five Robert Robot units were activated 67661 times during 28 hours of play activity over four days.

## 6.1 Study under Controlled Play Settings

We used descriptive statistics to assess the preliminary results of the questionnaires from our controlled play setting combined with the findings from logged data. Questions relating to situations not relevant for a specific respondent or play session (for example, questions relating to playing with a stranger when they had played with known mate) were excluded.

*6.1.1 Logged Data from the Controlled Play Setting*
The log files in the Robert Robot app supported analysis across all play sessions in the controlled play setting after confirming their average duration approximated 20 minutes. The most activated character in the test area over the course of the study was Robert Balloon with n = 3687 activations.

Using independent samples T-test we found that, on average, unfamiliar children (M = 95.50, SD = 5.02) activated Robert Robot more compared to familiar children (M = 75.8, SD = 28.16) per play session. While not a definitive reflection of actual play engagement, across the two groups, unfamiliar children put approximately 33.14% more bricks (M = 2233.25 grams, SD = 1108.62 grams) in Robert Robot's toy box, compared to familiar children (M = 1677.40 grams, SD = 809.64 grams). Based on the guardians' self-reported questionnaires, Robert Robot broke the ice between unfamiliar children to some extent (M = 3.18, SD = 1.13).

*6.1.2 Happiness and Fun Across Age Difference*
The analysis uncovered few statistically significant instances from which we highlight one compelling case to investigate further. We found that the robot made children with an age difference of less than two years happier (M = 4.33, SD = .65) than children with an age difference of two years (M = 3.67, SD = .62) (figure 4).

This is the only significant instance in this scope of our quantitative analysis, t(25) = 2.72, p < .012. Furthermore, children with an age difference of less than two years had more fun (M = 4.33, SD = .49) than children with an age difference of two year (M = 4.07, SD = .59) and children with an age difference of less than two years had more fun playing with each other (M = 3.45, SD = .82) than children with two years difference (M = 3.07, SD = .88).

Additionally, we found that unfamiliar children (M = 4.20, SD = .42) and children familiar with each other (M = 4.18, SD = .64) had an approximately equal amount of fun. On average, children who know each other agreed to a greater extent that they had fun with their playmates (M = 3.67, SD = .71) compared to children did not who know each other (M = 3.00, SD = .87).

*6.1.3 Differences in Same-gender and Mixed-gender Play*
Besides the age difference and affiliation, we saw that Robert Robot made both same-gender (M = 4.00, SD = .47) and mixed-gender groups (M = 3.94, SD = .83) equally happy. Furthermore, Robert Robot made it more fun to interact with the other child when same-gender children were playing together (M = 3.33, SD = .50) than two children with mixed-gender (M = 2.75, SD = .86), meaning that same-gender play was more fun than mixed-gender play.

*6.1.4 Assessing Socialization by Querying Guardians*
While not statistically significant, we saw indications that it was agreed to some extent that Robert Robot broke the ice between unfamiliar children (M = 3.18, SD = 1.13), and the unfamiliar children were slightly more engaged in the play (M = 4.18, SD = 0.73) compared to familiar children (M = 4.10, SD = 0.32).

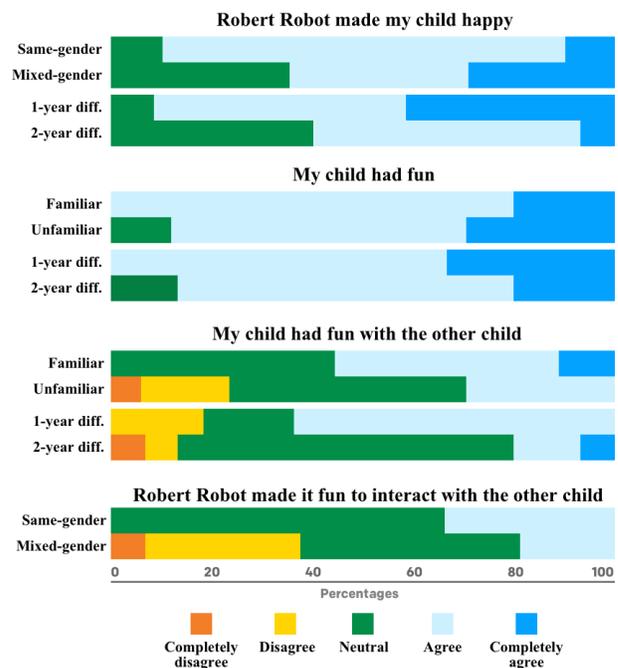

**Figure 4. Bar charts of select questions from questionnaire.**

Furthermore, our results show that the unfamiliar children usually play less actively (M = 2.59, SD = 1.06) which suggest that Robert Robot may have an impact on the physical play between unfamiliar children.

### 6.2 Findings in the Public Area

We now look at the results from the public area. This section details findings from the logged data before outlining our qualitative results. We observed a total of 212 play sessions, ranging from solitary play to parallel play, and group play across all constellations of play partners.

*6.2.1 Logged Data from the Public Area*

In total, the four Robert Robot units in the public area were activated 61078 times. Overall, Robert Balloon was the most active unit across all four days of the study, with 15979 activations in the public area. The least popular unit was the standard Robert Robot with 14344 activations. Collect requests were answered the most frequent. 70.21% of all collect requests (99 of 141) are responded to, while 65.68% (67 of 102) build requests are responded to. Interestingly, on average, collect requests are responded to in 11.29 seconds, while build requests, on average, are responded to in 11.34 seconds. Hence, there is no significant difference in response time between the two vastly different conceptual request types. We analyzed the response times between when a request was played and when they were responded to. The quickest response time was found to be the build request "*[Airplane sound effect] Will you try building an airplane for me?*" with an average response time of 5 seconds. The second fastest request responded to is "*Now we are in the middle of the ocean! Can you find some fish for me?*" with an average response time of 5.5 seconds.

*6.2.2 Qualitative Findings from the Play Observation Scale*

Table 1 and Table 2 detail observed play- and non-play behavior. We found that Robert Robot supports all the three social play types: solitary, parallel and group play. The most popular social play type is parallel play (60.85%), followed by group play (30.66%) and solitary play (8.49%). Of the cognitive play types, children engaged predominantly in exploration play (68.40%). Interestingly, construction play and games have somewhat limited representation in our findings, by only accounting for 4,72% of all play sessions respectively. We observed 78 non-play behavior events in the public area in which active conversation accounted for 53.85% of non-play behavior. Often, the social play had an audience of onlookers (23.08%) and unoccupied children (12.82%), who did not partake directly in social play.

### 6.3 Qualitative Observations in the Public Area

Across all three note takers, 382 sections of jotted notes were written. The duration of the play sessions ranged from around 20 seconds to approximately an hour, with most of the sessions lasting less than 20 minutes. Interview data and observation notes were analyzed using open coding to extract three themes from the data [42,43].

### 6.4 Children Broadcast Themselves and Mimic Each Other

We observed that Robert Robot breaks the ice for unfamiliar children and keeps them in a fun, physically active play loop. Two phenomena in particular help in breaking the ice, (1) the fact that children repeat what the robot says, and talk directly back at the robot, as a way of broadcasting themselves to their peers, and (2) that children mimic each other's play patterns to initialize social play (figure 5). Findings relating to the two phenomena will be elaborated on in depth.

*6.4.1 Repeating Robert Robot and Relaying Audio Cues*

The first communication between the children involves, in a majority of instances, the children repeating and relaying something Robert Robot has said. For instance, Robert Robot might say: "*Can you find some blue bricks for my toy box?*" and one of the children might respond: "*He needs blue bricks!*" which will then prompt both children to look for blue bricks. This soon turns into cooperation and exchanging of words among the children. All of this happens before the children have been formally introduced or approached one another directly, and the simple request from Robert Robot turns into an ice breaker between the children.

The children reacted intensely to auditory cues, especially simple words, funnily articulated phrases, and juvenile sound effects. The children share laughter and eye contact with each other and Robert Robot when they found something funny. The audio served multiple purposes; it made the children focused, attentive and we observed how it engages them more and is more actively shared among the children in the form of repeated voice lines. It is used as a method of broadcasting themselves to an audience, for example by saying: "*Haha, he said it smelled funny!*".

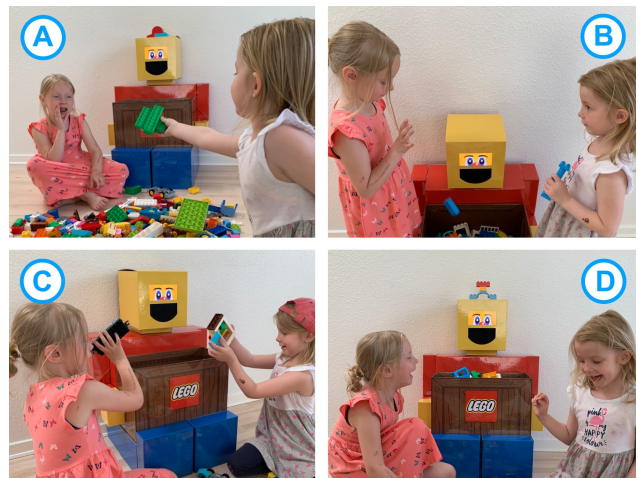

**Figure 5. Four common social play scenarios. A: One child relaying a request to another child. B: One child mimicking another child. C: Children putting random bricks in the toy box. D: Children sharing laughter.**

| Play Behavior | Frequency (N = 212) | Distribution (in %) | Exploration (N = 145) | Functional (N = 34) | Construction (N = 10) | Dramatic (N = 13) | Games (N = 10) |
|---|---|---|---|---|---|---|---|
| Solitary Play | 18 | 8.49 | 11 | 2 | 0 | 1 | 4 |
| Parallel Play | 129 | 60.85 | 92 | 21 | 8 | 3 | 5 |
| Group Play | 65 | 30.66 | 42 | 11 | 2 | 9 | 1 |

**Table 1. The table outlines quantified play behaviors observed in the public area.**

Humor appears to be an important factor in this regard too, as the children would often laugh and comment on funny voice lines or sound effects, like fart sounds. Sounds helped articulate the openness of the play and allowed the children not to be visually focused on the robot, but on all aspects of the play; their play partner, the bricks, and their guardians.

### 6.4.2 Mimicking Others and Seeking Praise

We saw countless instances of mimicry between both unfamiliar and familiar playmates [13]. The mimicry relates to both mimicking another child's play patterns (including learning how to activate Robert Robot in the first place) and mimicking affection given to Robert Robot, like kissing or hugging him. After mimicking another child's play, the children would often seek recognition ("*Look how many bricks I'm tossing in!*") from peers and guardians, which was often met with cheerful encouragement and applause. This loop would often repeat; having gotten praise for doing something "correct" (putting bricks in Robert Robot's toy box as requested), the children would continue doing so for the duration of the play. Often, this would evolve into competition-like play when children would be "showing off" by being very physical and repeatedly throw bricks, sometimes from, akin to basketball.

### 6.4.3 Age Differences Affects the Mimicry and Physical Play

In general, the older child takes on a mentoring role; they often eagerly demonstrate Robert Robot to their younger peers, who then mimic their "mentor". The older age group of children also expand the physical play to a more extensive physical area; they are more willing to move longer distances to collect bricks, and they are, in general, much more physically active in the play. The play of the younger age group is much more focused on the immediate surroundings around Robert Robot, often involving their guardians.

### 6.5 Responding to Robert Robot's Requests

#### 6.5.1 Collect, Build and Pretend Requests

It is evident that collect requests initiated the most activity with 26712 activations compared to the two other types of requests (build request with 17908 activations and pretend with 16458 activations). From the data logs, we found that collect requests are answered the fastest with a mean value across all three characters being 9.77 seconds compared to build requests where children used 14.62 seconds. This is indicative of a trend we observed throughout the study: often, the specific request from Robert Robot did not matter – the children would simply put in whatever bricks were nearby to hear and see Robert Robot's reaction. In addition to this, the children across age groups and affiliation hardly ever built something together before giving it to Robert Robot. We expected this when designing Robert Robot, which is why the underlying system is designed to play collect requests more frequent than build requests, but we were surprised by just how diligently children ignored the build requests. Rather, the children would often help each other in collecting bricks and await Robert Robot's response together. Prompted by this finding, we analyzed the data logs to see if there was a tendency for build requests to be fulfilled more often towards the end of a play session when the play would presumably have transformed into group play. There is no significant sign that this is the case. Additionally, we did not observe whether the children's age affected their tendency to respond more correctly to Robert's requests, collect requests, or otherwise.

| Non-Play Behavior | Frequency (N = 78) | Distribution (in %) |
|---|---|---|
| Unoccupied | 10 | 12.82 |
| Transition | 5 | 6.41 |
| Onlooker | 18 | 23.08 |
| Active conversation | 42 | 53.85 |

**Table 2. Observed non-play behaviors in the public area.**

#### 6.5.2 Collect Requests Prompts the Most Social Play

We found that collect requests resulted in faster response times and more requests fulfilled than the two other request types (56.29% of the total amount of fulfilled requests). This might be because collect requests are better at facilitating solitary or parallel play instantly and spontaneously, which can then evolve into group play over time. Building something together requires collaboration, coordination, and simple rules [1], which are not defined at the beginning of the play, which might explain why collect requests are better at activating the children – the play when collecting is simply simpler than building. Initially, the children might respond diligently (fx. give Robert Robot a red brick when prompted to), but the play soon develops into simply throwing bricks into the box to trigger a response from Robert Robot. This kind of aleatory play resembled that of interactions with one- armed bandit slot machines; the children would put something in Robert Robot's toy box, Robert Robot would react with a voice line and eye gaze animation, and that loop would repeat. The instant gratification of their actions seems to keep the children in an active play loop, both in terms of solitary, parallel, and group play. In these instances, the children would hardly ever build a creation to put in the box. Instead, they would take whatever bricks were nearby and throw them in. Additionally, a common observation was that the build requests sometimes became challenging for the children, especially "*Oh no, all my balloons have flown away! Can you build some new balloons for me?*". The play would frequently pause completely following this request. The request might be too intangible or age- inappropriate explaining why some children might not know how to react. Occurrences like this one had little consequence in the public area where the children would often merely throw any random surrounding bricks in the toy box.

### 6.6 Initial Interactions Can Be Both Enchanting and Frustrating

*6.6.1 Robert Robot's System Design Results in Both Enchantment and Frustration*
Children vigorously explored different techniques for eliciting an audio cue from Robert Robot because it is not immediately apparent what activates Robert Robot. A side effect of this unclarity is that the initial interaction with Robert Robot is often either rather enchanting or frustrating. Robert Robot's design allows for serendipitous moments of interactions between the children and the robot because the very first interaction might startle or surprise the children in a way that makes it seem like Robert Robot senses the children to an even more advanced and personal degree than it really does. In actuality, the event is entirely random.

Frustration often arises when Robert Robot sends out a request, and the children would spend a long time fulfilling the request. The system is designed to maintain a steady flow of requests going to avoid that the children become disinterested. In instances where the children were immersed in the play and already fulfilling a request, new requests would continually be said out loud, actively leaving the children playing "behind", sometimes discouraged from not being able to satisfy Robert Robots request immediately.

## 7 DISCUSSION
This section discusses three themes: *socialization between strangers*, *breaking the ice between preschool children*, and *the consequences of the simplified robot design*, before presenting four implications for design.

### 7.1 Socialization Between Children
This exploratory study highlights varying ways in which Robert Robot supports socialization between familiar and unfamiliar children. In the present study, the majority of all social play observed was parallel play (60.85%). One might argue that the closer the play approaches group play, the more socialization is occurring. Thus, striving towards supporting or encouraging group play is one of the primary goals for socialization. The authors regard *socialization* as the social interactions occurring across social play types outlined with the Play Observation Scale [38], and thus, the more the play resembles or approaches group play, the higher the socialization between the children.

Though group play may be the zenith of the social play types, parallel play can be equally socially rewarding for the children to partake in; it is merely a different type of social play. A possible explanation why parallel play predominantly was observed as outlined stems from the play context; a shared play area on an exposition floor with thousands of guests, most of whom are unfamiliar and have never interacted before. A follow-up consequence of this is the limited duration of the play; a short play session may only unlock the initial, spontaneous short-term engagement with an unfamiliar play partner. Parallel or group play would sometimes regress to solitary play, most often in situations where one child would eventually focus on building something by themselves, or after losing interesting in the play, after which the interaction with Robert Robot stops. It is interesting to note that Robert Robot does support solitary play, which speaks to the flexibility of the design; though group play might be encouraged for socialization, individuals playing solitarily with Robert Robot can have a fun experience. A fundamental finding is that any social play is occurring at all [7,32]. When children are interacting or communicating, be that by repeating Robert Robot, sharing their thoughts and exclamations, or sharing ideas for the play, the children are benefitting from simply being with other peers, which is foundational when developing an understanding of said peers and the breeding ground of meaningful relationships [25,48].

As previously noted, guardians take on varying roles in the play, from being the play enabler (e.g. by introducing their children to Robert Robot) or play inhibitor (e.g. by physically removing them from the play). We observed how the guardians engaging in play with their children had a cascading effect because the guardians would often initialize conversation with other guardians. In a few instances, situations like these lasted hours while the play between the unfamiliar children transformed. Hence, socialization is not only present among children but their guardians as well.

Though the study utilized a Danish-speaking version of Robert Robot, clear indications that play with Robert Robot can engage children across cultures in collaboration was observed. Robots inherently are of no cultural background or race, and thus, children with different cultural backgrounds might come together around the play with no predisposed notion of race or cultural background to influence the play.

When the children become fully immersed in a collaborative build or play, it seems that the direct interaction with Robert Robot stops. It is as if the focus on the robot reaches a halt and shifts to the accompanying unfamiliar child; the social play between the children succeeds, while the robot recedes and becomes a secondary, passive play opportunity in the background.

## 7.2 Breaking the Ice Between Preschool Children

The idiomatic ice breaks once the initial engagement between two unfamiliar children becomes social play. Having a shared non-human companion appears to alleviate some of the initial pressure to make the first contact between unfamiliar playmates in a way that is fun and engaging. Two primary findings aided in initializing the socialization between unfamiliar children, namely repeating what Robert Robot says to initialize exchanging of words, laughter, and other gestures [37,41], in addition to mimicry being used to create a shared play experience. In both the public area and the test area, we observed how one aspect of the interaction breaking the ice occurs during the very first moments of the interaction, namely that Robert Robot *introduces itself* to the children. Robert Robot is actively acquainting with the children by sharing some information about itself – its name and love for collecting bricks – and then invites the children to share their names. The children who chose to do so would comply promptly and say their names out loud. This circumvented the need for the children to actively engage with one another during the first moments of their first meeting. The robot became a conduit for sharing basic information about oneself, and the children learn each other's names (if they choose to share them), without ever having to introduce themselves to each other actively. Once they know each other's names, they progress from being total strangers to having at least some basic personal knowledge about the other child, which aids the transformation of non-play into play.

### 7.2.1 Robert Robot's Support for Pretend Play Is Hit-or-Miss

Robert Robot is intended to support pretend play – what Rubin categorizes as dramatic play – through voice lines structured around imaginative tasks and play scenarios. Building on Vygotsky's work, Nicolopoulou et al. highlight pretend play as being particularly suited for the development of social competence, including cooperation [31]. In the case of Robert Robot, in most instances, we did not find the intended pretend play facilitation to be fully understood by the children, and thus, pretend play rarely contributed to socialization between the children. The current design cannot measure pretend requests being responded to by the children in any other way than registering an activation event. Therefore, the data logs alone cannot provide firm results in understanding the impact of pretend requests on socialization. We saw mixed results during our observations when some children would respond to pretend requests like "*Clowns just love doing tricks, like standing on one leg – can you stand on one leg?*" successfully while simultaneously being unable to comprehend that Captain Brick's "treasure chest" is the toy box. This incongruity was less prominent among older preschool children, but the pretend aspect of Robert Robot was mostly disregarded, nonetheless. Whether this is indicative of a more significant challenge for social robots to support children's pretend play, a failing on our part as observers of the play or simply a challenge with aspects of Robert Robot's design is uncertain.

## 7.3 Consequences of the Simplified Robot Design

A consequence of the simplicity of the design is that Robert Robot is purely one-way conversational. Respondents in both the test area, as well as children observed in public, treated the Robert Robot units as interlocutors, mirroring findings by Breazeal et al. [10]. Children would often talk to or shout at Robert Robot by directly answering his requests out loud and expect a reply. This had two effects; it allowed the children to broadcast themselves to their surrounding audience (like mentioned above), and it would often frustrate the children that Robert Robot would not reply. The children attributed human natural language skills to Robert Robot because it *sounds* human and not distinctly robotic. Thus, they presume they can converse with it like with any other human. When the robot does not reply intelligently, the children are frustrated. An interaction scenario between Robert Robot and a child might look like:

> RR: "*Can you build a cool hat for me?*"
> Child: "*I have built a hat for you! Twice!! What else do you want me to build?*"
> RR: "*Wanna pretend that we can fly like birds?*"
> Child: "*Hello! What else do you want me to build!?*"

The exchange becomes incoherent because it is one-way conversational, underlining the fact that the robot does not understand the directed conversation from the child. Examples like the one outlined would often be accompanied with banging on the side of Robert Robot's head or poking his eyes.

### 7.3.1 Enchantment and Frustration

When designing *social robots*, Breazeal highlights the importance of shared *understanding* that should be established between the social robot and its users, noting the necessity to relate and empathize with the robot [9]. This relation with Robert Robot is established either successfully or unsuccessfully with the very first engagements the children have with the robot. With Robert Robot's weak artificial intelligence [39] being based on randomness, the timing proved to be very important, as it affects whether the very first interaction between the child and the robot will be successful or not. The randomness of Robert Robot's system design resulted in both enchantment and frustration for the children. When serendipitously encountering Robert Robot right as he says "*Hi! My name is Robert Robot*", the child often assume that he intelligently noticed them, which, in many cases, simply enchanted the children. The children would respond to Robert Robot's approach with glee and eagerness. The interaction gives them an impression of Robert Robot having a degree of agency which is much more capable than it is. Conversely, when children's first encounters with Robert Robot did not result in instant recognition by the robot, the children would believe they were being ignored, misunderstood or misinterpreted, and they would in turn often respond with frustration, yell at the robot, or hit it repeatedly.

While not based on randomness, Robert Robot has weak artificial intelligence similar to the simple ELIZA program [49], a system which adapted the role of a person, and created the illusion of being socially intelligent by picking up on elements of a conversation but without an underlying framework for contextualizing events. The simplicity of the design of Robert Robot's system provides an illusion of agency, which facilitates levels of engagement with the children that more technically advanced social robots also achieve [5], even though the possibly unfortunate duality of the first engagement challenges the emergence of empathy. Even when it falls short, the failing of Robert Robot is often innocuous because the play loop simply goes on.

## 7.4 Implications for Design for Socialization and Play

Based on our findings, we define four implications for design [55] of robots for socialization and play.

*7.4.1 Adjust the Experience to the Needs of the Age Group*

A 5-year-old has completely different developmental needs than a 3-year-old [1], and how they connect with the robot and with each other can be vastly different. Consider tailoring the experience to varying needs of the children's developmental stage, including the tasks given to the children, the wording of sentences said by the robot and how the robot responds auditorily and visually to communication from the children (as outlined in section 7.3).

*7.4.2 Keep Tasks Suggested by the Robot Simple*

This extends the former implication. We noted how even moderately challenging wording confused the children across age groups. Many children even had no conception of what Robert Robot's "toy box" is. This makes sense, since the vocabulary and imaginative skills of the preschoolers are not yet fully developed. Still, we observed how even simple phrasing, which demanded children used their imagination, often fell short. "*Can you build a balloon?*" is word-for-word not necessarily challenging for a preschool child to process, but the underlying imaginative concept can be.

*7.4.3 Support Children's Need to Broadcast Themselves*

Simple questions like: "*What is your name?*" or "*What is your favorite color?*" can support the children in personal expression by broadcasting a piece of themselves to a broader group of peers. The overwhelming majority of the guardians suggested having the robot learn and remember the children's names throughout the play session. According to their feedback, this would create a stronger relationship between the individual child and the robot, and would likely increase the children's desire to engage with the robot long-term [23].

*7.4.4 Incorporate Universally Understandable Audio Cues*

The fact that robots can produce sound effects adds to the playfulness, which might better enable cooperation [16]. Sounds have the innate quality of being understandable with varying language skills. A focus on universally understood sounds and sound effects can engage socialization across age groups, gender and cultural backgrounds because sounds can be easily recognizable. Possible play scenarios might leverage audio cues which demand participation across a group; one child might listen for the audio and relay to an audience of awaiting peers.

## 8 RESEARCH CONTRIBUTIONS

With this study, we extend on the research in the field of cHRI by showcasing how a simplified robot can facilitate the first encounters between unfamiliar and familiar preschool children and engage them in social play. The product of our work is Robert Robot, an interactive artifact [54] in the form of a simplified robot comprised of a physical exterior accompanied by a custom smartphone application. Furthermore, we add to the body of work in cHRI with an empirical research contribution [54], with findings from our evaluation of Robert Robot with 30 preschool children and observations from 212 play sessions in a public environment, analyzed based on The Play Observation Scale [38]. Our subsequent findings informed four implications for design of robots for socialization and play.

## 9 LIMITATIONS AND FUTURE WORK

Our study is limited by the fact that we did not have equal distributions or sample sizes of unfamiliar and familiar children across all play sessions in the controlled play setting, which affects our results; this limited the chance of discovering significance in our findings. Larger, equal distributions and sample sizes would have made it easier to make more general assumptions about the play experiences preschoolers have when interacting with Robert Robot. As an adaptable research platform, avenues for future work include researching the effect of Robert Robot's anthropomorphism and exploring possibilities for sustaining children's engagement over time.

## 10 CONCLUSION

In this exploratory study, a collaboration with The LEGO Group, we found that Robert Robot is fun for both unfamiliar and familiar children. Robert Robot supports socialization among preschool children engaging in parallel, group and solitary play. Additionally, we found that Robert Robot breaks the ice between unfamiliar children by facilitating repetition and commenting on Robert Robot's audio cues. Furthermore, unfamiliar and familiar children mimic each other to initialize social play. The system design, built on randomness, resulted in both enchantment and frustration among the children.

The study contributes to the child-robot interaction research field with (a) an artifact contribution, Robert Robot, a simplified anthropomorphic social robot, and (b) empirical research contributions from a two-part study of Robert Robot in use by 30 preschool children in a controlled play setting, and 212 play sessions in the wild, resulting in four implications for design of robots for socialization and play.